\documentclass[preprint]{aastex}

\begin{document}
\title{Planet Formation by Concurrent Collapse}
\author{Michael Wilkinson$^{1}$ and  Bernhard Mehlig$^{2}$}
\affil{ $^{1}$Department of Mathematics and Statistics, The Open
University, Walton Hall,
Milton Keynes, MK7 6AA, England \\
$^{2}$Department of Physics, G\"oteborg University, 41296
Gothenburg, Sweden \\}

\begin{abstract}
After reviewing the difficulties faced by the conventional theory
of planet formation (based upon the aggregation of microscopic dust
particles), we describe an alternative hypothesis. We propose that planets form by gravitational collapse at the
same time as the star about which they orbit. This {\sl concurrent
collapse} hypothesis avoids theoretical difficulties associated with the conventional model and
suggests satisfying explanations for various poorly understood phenomena. We introduce new explanations for FU Orionis outbursts seen
in young stars, the discovery of exoplanets with eccentric orbits and the existence of small rocky objects such as chondrules
in the solar system.
\end{abstract}

\maketitle


\section{Introduction}
\label{sec: 1}

There is a consensus that the initial stage in planet formation involves aggregation of dust particles in the gas around a young star, with gravitational forces taking over once the aggregation process has built significantly larger objects \citep{Saf69}. It is believed that this gas usually forms an accretion disc. The facts that the solar planets have orbits which are roughly circular and coplanar with the Sun's equator are readily explained by the hypothesis that they formed in a disc.
However, there are some difficulties with this model, which we believe are sufficiently telling to compel a search for alternative mechanisms. In this paper we first summarise the difficulties with the standard model (which is reviewed by \citet{Pap+06}), before describing the consequences of our alternative hypothesis, which we outline below.

A star forms when a cloud of interstellar gas satisfies conditions which allow it to undergo gravitational collapse. We argue that as the star forms, the gas cloud fragments. Other regions of the cloud in the vicinity of the nascent star will condense by gravitational attraction: we term this process {\sl concurrent collapse}. Consequently, gravitationally self-bound objects which are much smaller than the star itself may be bound in the gravitational potential of the star. We term these objects {\sl juvenile planets} and we propose that the planets seen in mature planetary systems result from the interaction of these objects with the circumstellar accretion disc, and possibly with each other. The structure of the juvenile planets may be radically transformed by processes such as further accretion or ablation of material as they pass through the gas, or by collisions. These processes
could dramatically change the mass, orbital parameters, composition and number of the planets.

We remark that a related, but less radical suggestion has been made by \cite{Rib+07}, who propose that the largest extra-solar planets (exoplanets) may have formed by direct gravitational collapse. They point out that this mechanism can explain the large masses and high eccentricities of many of the recently discovered exoplanets \citep{But+06}. By contrast, we propose that the concurrent collapse mechanism could be the origin of all planetary systems. Other theories involving direct gravitational collapse of planets have been proposed in the past: we mention these in section \ref{sec: 3}, contrasting them with our own hypothesis. The consequences of our new theory are uncertain, but we argue that it has the potential to resolve all of the difficulties with the standard dust-aggregation model. Because we hypothesise that planetary systems developed from very different initial conditions, quantitative modelling of the processes which we consider is an open-ended problem. However, preliminary quantitative estimates of the processes which we describe using  plausible initial data yield very encouraging results.

\section{Difficulties with the standard model}
\label{sec: 2}

\noindent 2.1 {\sl Observations on exoplanets}

The discovery of substantial numbers of extra-solar planets \citep{But+06} has produced observations which challenge the standard model.

Significant numbers of these exoplanets have large orbital eccentricity.
Various models have been proposed to account for this \citep{Zak+04}.
The most plausible of these is a slow-acting three-body instability resulting in a drift of orbital parameters, leading to a near-collision between two planets. This could cause escape of one planet and scattering of the other to an eccentric (and probably non-equatorial) orbit \citep{For+03}. It is as yet not clear whether the large proportion of exoplanets with eccentric orbits can be explained by this model. The model would suggest that large planets are less likely to be scattered into highly eccentric orbits. There seems, however, to be a positive correlation between eccentricity and mass \citep{Rib+07}. There are also examples of planets where the orbital plane is at a very large angle of inclination to the equator of the star: see for example \cite{Joh+08}.

Many of the exoplanets are found to have very high masses, which suggests that the circumstellar accretion disk would have to have a much higher mass than is usually assumed. A related difficulty is the observation of substantial numbers of exoplanets which are gas-giants orbiting close to their star, at radii where the temperature would be too high to allow them to form {\sl in situ}. It is inferred that these {\sl hot Jupiters} have drifted closer to their star after their formation, but the mechanism for this is not fully understood. It has been suggested that the inward migration of the orbits of hot Jupiters is related to the their gravitational interaction with a circumstellar disc \citep{Lin+96}. This also appears to require larger quantities of gas in the circumstellar disc than are usually assumed. Another difficulty is that the timescale for planet formation by dust aggregation is expected to be slower at a greater distance from the star: the inner part of the circumstellar disc may no longer be present by the time a planet has formed at larger radius.

In general, the observations on extra-solar planetary systems show that these are very diverse. It is very hard to reconcile this diversity with a picture in which planets are all created under very similar and highly ordered circumstances in circumstellar accretion discs.

\noindent 2.2 {\sl Theoretical difficulties with the standard model}

There are theoretical difficulties related to the inferred existence of turbulence in circumstellar accretion discs. Evidence from surveys of young stars indicates that the accretion discs are relatively short lived, with lifetimes of order $10^6\,{\rm years}$ \citep{Pap+06}. The transport of angular momentum by viscous drag in a laminar flow is much too slow to allow the accretion process to proceed on this timescale. It is inferred that motion in the disc must be turbulent, allowing transport of angular momentum by turbulent diffusion. This poses two problems.

First, what is the origin of the turbulence? A rotating disc is stable \citep{Ji+06} and there is no fully persuasive model for the onset of turbulence in circumstellar accretion discs. This is not a
critical difficulty because of the highly uncertain structure of the
accretion disc and the many different possible sources of hydrodynamic
instability, especially when magnetic fields may have an influence due to
the gas being partially ionised \citep{Gam96}.

A more telling difficulty which follows from the existence of turbulence is related to the fragility of clusters of dust particles in a turbulent environment. The dust particles adhere due to van der Waals forces and to a certain extent electrostatic forces. Because these binding forces are weak, aggregates of dust particles are very easily fragmented by collisions. Estimates of the relative velocity of particles colliding in a turbulent environment \citep{Vol+80,Meh+07} indicate that the collision speed increases with the size of the particles, so that there may be a maximum value for the size of a dust cluster which can be formed by aggregation in a turbulent environment. Recent estimates indicate that the maximum size that can be reached by clusters of dust particles appears to be very small for reasonable values of the parameters in a model for the protoplanetary accretion disc \citep{Wil+08}.

Even if it is believed that dust grains can aggregate in this highly turbulent environment, there is another theoretical difficulty. The gas in an accretion disc is partially supported by its pressure, so that its orbital velocity is slightly lower than the Keplerian value in the quasi-static state. A \lq rock' (more accurately, an aggregate of dust grains and possibly ices) which is entrained with the gas is not supported by the pressure and consequently slowly spirals in towards the star \citep{Wei77,Tak+02}. This effect is most pronounced for rocks with size comparable to the mean free path of the gas (typically $1\,{\rm cm}$ to $1\,{\rm m}$), and the timescale for spiralling in is of the order of $100\,{\rm yr}$ starting from an orbit at $1\,{\rm AU}$. Even under the most favourable assumptions about growth rates by aggregation, it is difficult to see how aggregates of dust particles can grow sufficiently rapidly to avoid spiralling in.

Two mechanisms have recently been proposed to resolve the difficulties caused by spiralling in. \cite{Kre+07} propose that a magnetohydrodynamic effect may create a pressure extremum at the {\sl snow line}, that is at the limiting radius where the circumstellar disc is cool enough for ice crystals to form. Another proposal \citep{Joh+07} appeals to the effects of large-scale eddies in the gas forming the circumstellar disc to make long-lived pressure extrema. It is not clear whether either of these effects is able to trap dust aggregates for sufficiently long periods. Also, effects associated with the snow line can only facilitate the growth of a planet at one radius, and further work is required to explain how other planets might form.

\noindent 2.3 {\sl Other observational puzzles}

Yet another source of difficulty is that there are observed phenomena which
are not persuasively explained by the standard model. One of these is the occurrence of outbursts of increased luminosity of young stars. These outbursts (referred to as {\sl FU Orionis outbursts} after their first
observed exemplar) can increase the luminosity by a large factor (typically three magnitudes, often substantially higher \citep{Bel+95}). Usually the onset is rapid (with a timescale of $1-10\ {\rm yr}$) and the decay back to normal activity levels is much slower (with a timescale of $10-100\ {\rm yr}$). There is no indication that the outbursts are periodic, but the observation records are not sufficiently long for any periodic or quasiperiodic behaviour to have become apparent. The diversity of the outburst phenomena appears to be a challenge to finding an explanation. Attempts have been made to model the outbursts as an instability of the circumstellar accretion disc \citep{Bel+95}. The proposed instability is associated with an increase of opacity due to ionisation of hydrogen, which requires very high temperatures. Also, the rapid rise of the outburst suggests that the instability propagates inwards, which requires rather specific assumptions about the structure of the disc. More recently, it has been suggested that the instability might be triggered by the presence of a large planet embedded within the disc \citep{Lod+04}. There is observational evidence for periodic fluctuations in some stars in the outburst phase consistent with an embedded planet \citep{Her+03}, but the planetary mass assumed by \cite{Lod+04} is so large as to suggest that this mechanism is not a general explanation.

A further observation which is hard to explain is the existence of
chondrules, small grains of material, of size typically $1\,{\rm mm}$,
which are found in meteorites \citep{Hew97}. These grains show evidence of having been heated to high temperatures, causing them to fuse to a glassy structure. It is believed that the heating occurred over a short period (otherwise the chondrules may have sublimated). The relatively short exposure of large quantities of material to high temperatures has no satisfactory explanation in the standard model. It has been proposed that shocks \citep{Hew97} or nebular lightning \citep{Gut+08} caused the heating of chondrules, but there no persuasive evidence that either mechanism is capable of explaining the sizes or the total mass of chondrules.

\section{Concurrent Collapse hypothesis}
\label{sec: 3}

\noindent 3.1 {\sl The hypothesis}

Given these difficulties with the existing theory it is desirable to find a model which circumvents these problems. We therefore propose an alternative hypothesis, which we term {\sl concurrent collapse}, which was outlined in the introduction.

We propose that juvenile planets form by direct gravitational collapse from fragments of the molecular cloud which forms their parent star. Once the accretion disc has formed, its rotation resists gravitational instability, unless the assumed gas density is implausibly large. We therefore assume that the planetary precursors are established as gravitationally bound objects before the nebula collapses into a circumstellar accretion disc. The juvenile planets subsequently interact with the circumstellar disc in a variety of ways, which we discuss later. \cite{Rib+07} proposed that very large gaseous exoplanets formed by fragmentation of a protostar as it collapses. Our hypothesis goes further: we propose that all planets could form as a result of gravitational collapse. \cite{Cam78} proposed that gas-giant planets may form by gravitational instability of a circumstellar disc. We remark that similar models have also been considered in more recent works \citep{Bos03,Ric+05}, and that \cite{Kui51} had previously proposed similar ideas to Cameron. Our hypothesis differs from that of Cameron in that we assume that the juvenile planets form at the same time as the protostar, and before the circumstellar nebula has formed itself into an accretion disc. The model proposed by Cameron faces a difficulty, in that due to rotation of the accretion disc centrifugal effects resist gravitational collapse, unless the density of gas in the disc is assumed to be very large. Most importantly, it does not help to explain the existence of large planets with eccentric orbits.

\noindent 3.2 {\sl Outline of consequences}

Our model provides a natural explanation for the existence of planets with eccentric and inclined orbits, because the juvenile planets are most likely to be formed with such orbits. We argue below that in some cases they will be entrained into the circumstellar accretion disc, leading to circular and equatorial orbits, whereas in other cases the interaction with the accretion disc will not be sufficiently strong to cause entrainment, and planets will remain in eccentric orbits, as is frequently observed for extrasolar examples.

Our hypothesis also leads to explanations for phenomena which are not persuasively explained within the framework of the standard model.
For example, it suggests an explanation for outbursts of increased
luminosity in young stars (FU Orionis outbursts). Arguments that these outbursts result from an instability of the accretion disc are not entirely persuasive, nor is there any evidence of a change which precedes the relatively sudden onset of the outburst. Within the framework of our hypothesis, the occurrence of many different types of outburst effects could be explained. One-off outburst events could be explained
by collisions between juvenile planets hidden within an accretion
system. If periodic outbursts are eventually observed, they could be
explained by a juvenile planet which orbits with a long period orbit (which is either eccentrical or out of the plane of the accretion disc)
disrupting the disc every time its orbit passes through.

Our proposal also suggests two distinct mechanisms for the production of chondrules. We argue below that our model may allow energetic collisions of juvenile planets, which could generate small fragments which are heated to high temperatures for a short period. Also, the passage of a body moving at high speed through the gas in the circumstellar accretion disc could result in frictional heating of its surface (from which liquid droplets could be ablated) or to the production of shock waves. (The possible role of shock waves in the production of chondrules was considered in \cite{Gut+08}, but in that work a source for the shock waves with sufficient energy was not identified).

Our concurrent collapse hypothesis avoids the difficulties in building up clusters of dust particles by aggregation and of small objects spiralling in to the star, because the planets are formed directly by gravitational collapse. It also provides an additional mechanism to explain the origin of turbulence in protoplanetary accretion systems, because the presence of large bodies moving through the gas will create turbulent motion.

The presence of large quantities of gas in the accretion system at a
time when large planets have already formed helps to explain the existence
of hot Jupiters. As well as the indirect mechanism involving gravitational interaction between the circumstellar disc and the planet described by \citet{Lin+96}, it is possible that displacement of the accretion disc gas by the planet could be the dominant effect, as illustrated by estimates in section 4.2 below. The interaction of a poorly condensed juvenile planet with the high density of gas in an immature accretion system could create sufficient drag to cause inward migration.

According to our hypothesis, all of the juvenile planets would form with an elemental composition representative of that of the interstellar medium. It is necessary to explain how small rocky planets or ice planets are formed as well as gas giants. If we assume (following \cite{Sla+80}) that the juvenile planets develop a rocky core, then two possibilities present themselves. First, it is possible that gas could be pulled away from a planet as it passes through the circumstellar disc, leaving a smaller rocky core. The other possibility is that after a collision of a pair juvenile planets, solid fragments will combine to form small rocky planets. Provided that a high proportion of the rocky debris is significantly above $1\,{\rm m}$ in size, the difficulties of rocks spiralling in to the star are avoided.

\section{Discussion}
\label{sec: 4}

Because our concurrent collapse hypothesis involves assumptions about the
conditions at the early stages of the life of the stellar accretion system, its implications for the eventual structure of a planetary system are uncertain, and will require substantial additional research. In this paper we can only give the first indications of how the problem might be approached. There are several questions which must be addressed. The first concerns whether the juvenile planets can indeed form directly by gravitational collapse. If so what can be said about their masses, their orbital parameters, and the extent to which the rest of the protoplanetary nebula has organised into an accretion disc by the time they have formed?
Then we must consider how the juvenile planets and their orbital parameters will be modified by interaction
with the accretion disc. Finally, are the juvenile planets likely to undergo collisions, and what would be the consequences of these collisions? We make preliminary and in many cases speculative remarks about these questions in the remaining sections of this paper. However, even at this stage we can advance quantitative arguments which support our hypothesis.

\noindent 4.1 {\sl Gravitational collapse}

First we consider the properties of the juvenile planets in their initial state. Stars form by gravitational collapse of a molecular gas cloud. The high masses of molecular clouds and the much smaller and relatively narrow range of initial stellar masses indicate that the molecular cloud fragments as it collapses, and that the processes that determine when fragmentation ceases are of crucial significance in determining the masses of stars.
We propose that the collapse and fragmentation process produces many smaller bodies, our \lq juvenile planets', which are formed concurrently with the nascent stars, and which are gravitationally bound to the protostars.

The process of gravitational collapse and fragmentation is not yet perfectly understood. One issue which has been perceived as a difficulty in the past is that first-principles estimates for the masses of objects produced at the final stage of fragmentation tend to be much smaller than typical stellar masses. \cite{Low+76} estimated a minimum mass for objects produced by gravitational collapse which is approximately one hundredth of the solar mass. Their calculation can be criticised at various levels, but it has not been possible to develop a first-principles estimate of the stellar mass which produces much higher estimates. Further work on this topic (discussed in \cite{Wil+08a}) has consistently yielded estimates which are considerably smaller than a typical stellar mass. There is an alternative way out of this difficulty which avoids having to propose that the fragment mass is significantly higher than the estimate in \cite{Low+76}. That is to assume that the initial masses of the final fragments are in fact comparable to those of gas-giant planets, and that the much larger masses of protostars are determined by the subsequent accretion of material. This alternative model receives support from numerical simulations of star formation \citep{Bat+05}, and is also supported by a model for gravitational collapse which we have elaborated elsewhere \citep{Wil+08a}.

In the following we adopt the hypothesis proposed by \cite{Bat+05} that the first gravitationally bound objects which are stable against fragmentation (we refer to these as \lq dense cores') have masses which are more comparable to gas-giant planets than to typical protostars. The production of protostars therefore depends upon the subsequent accretion of gas onto these objects.
It is natural to extend this \lq light core hypothesis' to assume that the rate and extent to which these dense cores can accrete additional material will vary widely, depending on their environment. We assume that some of the \lq dense cores' accrete material to become protostars but that others develop relatively slowly and remain of planetary mass. Many of these slowly accreting objects will become bound in the gravitational potential of the protostars. These are the juvenile planets of our theory.

There are two pieces of evidence which appear at first sight to contradict our hypothesis. We now address these in turn.

One argument which suggests that stars and planets have a different origin is the fact the distribution of masses of known astronomical bodies shows a pronounced minimum at a mass corresponding to brown dwarf stars. This \lq brown-dwarf desert' does not necessarily indicate that planets cannot be formed by gravitational collapse, in the same way as stars. After a dense core forms, in order to become a protostar it must accumulate a substantial amount of additional mass by accretion of material from the molecular cloud in which it formed. A juvenile planet, forming by gravitational collapse of a fragment of a nascent protostar, may not be able to accrete additional material, because of the competitive effect of the larger protostar. In fact, we give arguments below indicating how juvenile planets are expected to lose mass due to ablation. If, as seems likely, larger bodies accrete mass more rapidly, it is possible to have a bi-modal mass distribution even if both stars and planets are formed by gravitational collapse.

Another argument which is used to support the particle aggregation theory of planet formation is the observation that there is a correlation between the metallicity of stars and the probability of them hosting planets \citep{Fis+05}. This appears to be hard to reconcile with our hypothesis of direct gravitational collapse, because the \lq metals' are a small fraction of the mass of a molecular cloud. However, the mass of the dense cores produced by gravitational collapse is determined by the opacity of the gas in the molecular cloud, and we predict that the typical fragment mass is inversely proportional to the opacity \citep{Wil+08a}. The Wien wavelength corresponding to the typical temperature of a molecular cloud (of order $10\,{\rm K}$) is in the far-infrared range, where the opacity is dominated by dust. We may therefore assume that the opacity is proportional to metallicity. We therefore argue that high metallicity does in fact favour the production of small juvenile planets around a protostar, in agreement with the correlation noted by \cite{Fis+05}.

\noindent 4.2 {\sl Evolution of orbital parameters}

We have arrived at the view that a nascent star can be surrounded by both an accretion disc and one or more juvenile planets. There is no reason why the juvenile planets should have orbits which are circular, or coplanar with each other or with the accretion disc.

If the juvenile planets are not in circular orbits, or if they are not
in the plane of the protostellar accretion disc, they will be moving through the gas forming the disc with a substantial speed
(probably a significant fraction of the orbital speed; this might
be in excess of $10^4\,{\rm m\,s}^{-1}$). We now argue that this interaction may cause the juvenile planet to become entrained in the flow of the accretion disc. The criterion determining whether entrainment occurs is as follows. If the volume swept out by the trajectory of the juvenile planet contains a mass of gas which is large compared to its own mass, then most of its momentum is transferred to the gas in the protostellar accretion disc and it becomes entrained. Let us consider a quantitative estimate: we consider a case where the juvenile planet has collapsed rapidly to form a relatively dense body. We assume that the juvenile planet has density $\rho_{\rm p}\approx 10^3\,{\rm kg\,m}^{-3}$ and moves through an accretion disc in a roughly circular orbit at radius $R\approx 10^{12}\,{\rm m}$. The density of gas in the protostellar disc is quite uncertain and it must be highly variable; we assume that the gas density in the accretion disc at this radius is $\rho_{\rm g}\approx 10^{-6}\,{\rm kg\,m}^{-3}$ (this is a representative figure, consistent with the data used by \citet{Wil+08}). Finally, we assume the juvenile planet has mass $M_{\rm p}=10^{27}\,{\rm kg}$ (approximately $10^{-3}$ solar masses), corresponding to a linear dimension $a\approx 10^8\,{\rm m}$. If the orbit lies within the disc, with each orbit the juvenile planet therefore displaces a mass of gas approximately equal to $10^{23}\,{\rm kg}$. This estimate indicates that it will be entrained by the accretion disc over $10^4$ orbits, that is within a period of approximately $10^5\,{\rm yr}$. If the lifetime of the protostellar accretion disc is approximately $T=10^6\,{\rm yr}$, in this case we conclude that the juvenile planet would reach a nearly circular and equatorial orbit, after making a very large number of eccentric orbits.

The values of the mass of the juvenile planet and the density of gas in the accretion disc could vary by orders of magnitude. In one extreme, where the juvenile planet has not yet collapsed to a high density and moves through an accretion disc with higher density than was assumed above, its motion may be entrained to the disc within the period of a few orbits. For a the situation described by the estimate above, a juvenile planet may make a large number of orbits before becoming entrained to the disc. In other cases, where the density of gas in the disc is low or when the juvenile planet has an inclined orbit (which spends little time passing through the accretion disc), it might not be entrained by the disc at all. The model therefore predicts that some planetary systems would end up with planets in inclined and eccentric orbits, while others would have all planets in roughly circular and equatorial orbits. In line with the arguments presented above, the current list of extrasolar planets contains many examples which have highly eccentric orbits, as well as examples with near circular orbits \citep{But+06}.

Most discussions of the interaction of planets with gaseous matter in the accretion disc have focused on their gravitational interaction: the planet causes a gravitational disturbance of the disc, and the resulting gravitational field from the disc can cause the planet to spiral in \citep{Lin+96}. This model is appropriate when the planet moves in an approximately circular orbit, at nearly the same velocity as the gas in the circumstellar disc. However our model allows for planets to be created in eccentric orbits and non-equatorial orbits, so that their orbits will involve large velocities relative to the gas in the accretion disc. In our model different physical processes are relevant. We have shown above that direct momentum transfer is important, whereas the larger relative velocity is expected to reduce the effectiveness of the gravitational interaction discussed by \cite{Lin+96}.

\noindent 4.3 {\sl Evolution of the structure of juvenile planets}

The gravitational collapse of the juvenile planets proceeds until it is balanced by the effects of pressure and centrifugal forces. The latter are  expected to determine that the initial state of a juvenile planet will itself be in the form of a disc. The rate at which it can collapse by expelling angular momentum are unknown. Also, random variations of the angular velocity of the core of the collapsing gas cloud will result in a very wide variation of the initial density of the juvenile planets.

It might be expected that the juvenile planets will contain a dense core of solid material, resulting from the dust (approximately 1\% by mass) sinking to the centre, but this is far from certain. It is difficult to estimate the timescale for dust particles to settle to the centre, because settling may be resisted by turbulence or facilitated by effects associated with convection. \cite{Sla+80} proposed a model for the formation of dense cores inside giant planets (which they applied to the model in \citep{Cam78}): they argued that precipitation of liquified dust particles could be an efficient route to forming a rocky core.

As noted earlier, the juvenile planets may move through the accretion disc with high velocity. If the juvenile planet has sufficient angular momentum about its centre of mass, its initial state will be a disc of gas with a low density. The outer gaseous disc may have an escape velocity which is comparable with the speed at which the the juvenile planet passes through the accretion disc. This weakly bound gas could easily be ablated from the juvenile planet, reducing its mass to the point where lighter elements will evaporate.

Thus we argue that the structure of the juvenile planets can change in a variety of ways. Rocky material may precipitate into their cores. The mass of the planet could be increased by sweeping up gas from the accretion disc, or reduced by ablation. If dense material can sink to the core of the planet, the latter mechanism has the potential to produce relatively small rocky planets from a large juvenile planet with an initial composition representative of the interstellar medium.

The motion of a planet at high speed through an accretion disc will cause frictional heating, which could be related to the mechanism for formation of chondrules: we conclude this section by discussing two different scenarios.

The passage of a juvenile planet at high speed through the accretion disc could generate shock waves and high temperatures. It has been proposed that shock waves might be responsible for the high temperatures which are necessary to form chondrules \citep{Gut+08}, but the known sources of shock waves do not appear to provide sufficient energy. Our model provides a source for shock waves which could heat very large masses of material to sufficiently high temperatures. For example, if a planet moves at a speed of $v\sim 3\times 10^3\,{\rm ms}^{-1}$ relative to the accretion disc, this will create a shock wave in which gas is heated to temperatures of order $10^3 {\rm K}$, and some of this gas would be trapped behind the shock wave for a time or order $a/v$, where $a$ is the radius of the planet (this is of the order of a few hours). Also, in section 4.2 it was demonstrated that a mass of gas comparable to the mass of the planet will eventually be displaced. Our model therefore makes the suggestion that shock waves may be responsible for the formation of chondrules more plausible, but further work would be required to determine whether the sizes of chondrules can also be explained.

An alternative scenario for the formation of chondrules is that a juvenile planet with a rocky core has all of its gaseous material ablated away. The surface of the planet will be heated to a similar temperature to that considered above,  high enough to melt the surface of the planet. The very high speed of the gas over the surface could lift \lq spray', in the same fashion as water droplets are lifted by wind from the surface of the sea. These droplets would solidify on cooling, becoming solid chondrules distributed throughout the accretion disc.
Neither of these scenarios can readily be dismissed, but substantial additional work is required to make a quantitative theory. A further mechanism for chondrule formation is discussed in section 5.2 below.

\section{Implications of collisions}
\label{sec: 5}

The juvenile planets are expected to be created in eccentric and inclined orbits. If there is more than one such planet, this raises the possibility that there will be either collisions, or very close encounters which cause the tidal disruption of one or both objects. Scattering events which result in a significant change of orbital parameters (and possibly ejection of one of the planets) may also occur.

First we must ask whether the probability of collision can be significant. We consider a quantitative estimate: let us assume that the orbital radii of two juvenile planets are approximately $1\,{\rm AU}$, so that the area of the plane occupied by the orbit of one of the juvenile planets is approximately $10^{22}\,{\rm m}^2$. Further we assume that the cross-sectional area of a large planet with high density is $10^{16}\,{\rm m}^2$, so that the probability of a collision with each pass is approximately $10^{-6}$. These figures suggest a collision rate of one collision every $10^6\,{\rm yr}$, that is one collision in the life of the accretion disc. By changing the assumed data to other equally plausible values, the probability of collision within the lifetime of the circumstellar disc could be changed dramatically, from collisions being highly improbable to almost certain. For example, if it is assumed that the juvenile planets spend a long time as diffuse discs rather than compact spheres, the probability of collision is greatly increased.

We conclude that of collisions of juvenile planets are a possibility which must be considered. We discuss two possible consequences of such collisions below.

\noindent 5.1 {\sl Collisions as an explanation for outbursts}

We argued above that it is possible for juvenile planets to collide, because they may have different eccentric orbits. The collision velocity could
easily be comparable to the orbital speed. This could give a collision
velocity of $10^4\,{\rm m\,s}^{-1}$, and even higher figures are possible. Such energetic collisions could cause fragmentation of planets with the debris heated to very high temperatures.

Collisions between juvenile planets could be an explanation for
the outbursts of increased luminosity which are observed in some
young stars (FU Orionis outbursts). These typically have a relatively rapid onset and a much slower decay. We argue that the timescales of onset and decay of outbursts are consistent with them resulting from an increased rate of accretion onto the protostar following a collision. First, larger pieces of collision debris could be scattered into orbits which take the fragments directly into to star. Secondly, the collision might be expected to  also produce smaller fragments which would be slowed by motion through the gas in the circumstellar disc, and which would spiral into to star slowly by the mechanism described by \cite{Wei77}. We note that the typical rise time of the outbursts ($1-10$ years) is consistent with the former process, whereas the decay time ($10-100$ years) is consistent with the timescale for smaller objects spiralling into the star.

We can also estimate the increase in luminosity due to this mechanism, which is due to material falling into the star.
The mass of material which falls into the star will be very variable, depending upon the collision parameters as well as the masses of the colliding objects. This is consistent with the very wide range of luminosities of outbursts. Let us estimate the luminosity of an extremely intense outburst, caused by a mass of material
comparable with the mass of a large planet ($10^{-3}$ solar masses, say) spiralling into the star over a period of perhaps $10\,{\rm yrs}$. This accretion rate exceeds the commonly assumed baseline luminosity resulting from an accretion rate of $10^{-7}$ solar masses per year \citep{Pap+06} by a factor of $10^3$. This is sufficient to explain the most pronounced outburst events, which involve an increase of luminosity of up to six magnitudes. In most cases the collision will scatter only a small fraction of the planetary mass into the star, which is consistent with the fact that FU Orionis outbursts are less intense that this estmate, and vary widely in magnitude.

We conclude that a mechanism based on collisions between juvenile planets is capable of explaining the order of magnitude of the timescales and strengths of FU Orionis outburst events, as well as giving an insight into the diversity of these phenomena.

\noindent 5.2 {\sl Collisions as a route to producing rocky planets and chondrules}

In section 4.3 we suggested that small rocky planets could be produced by ablation of light elements from a juvenile planet, leaving behind a rocky core. We also argued that the high temperatures required to produce chondrules could be produced by frictional heating due a planet moving through the accretion disc at high speed. Here we propose another mechanism for producing rocky planets and chondrules. If the juvenile planets form rocky cores, and two of them were to undergo a collision, rocky debris would be widely scattered. In the following we discuss whether this debris could coalesce to form rocky planets, and whether the debris could include particles resembling chondrules.

The most severe difficulty with the standard model for planet formation lies in the fragility of aggregates of dust. If the circumstellar disc contained larger pieces of rocky material, which would be much less easily fragmented by collisions and which would avoid spiralling in, the standard model would be much more tenable. In particular, the largest fragments would play the role of the planetisimals of the standard theory, and rocky planets could be formed by coalescence of these bodies and by sweeping up smaller bodies, as is envisaged in the standard model. Most of the difficulties of growing large bodies are avoided, and rocky planets on near-circular and coplanar orbits could be formed from collision debris.

The collision of two juvenile planets with rocky cores would be a violent event, generating very high temperatures. We argue that the debris could include very small droplets of molten material which would solidify to smooth walled and roughly spherical glassy grains, resembling chondrules. Let us now estimate the size of droplets. The droplet is stable against fragmentation due to shearing provided the ratio of the internal kinetic energy of the fluid to the energy associated with the surface tension is small (in fluid dynamics this ratio is termed the Weber number, and is used to characterise the sizes of rain droplets). If the typical velocity of relative internal motion is $v$, and the size of the drop is $r$, the kinetic energy is $E_{\rm kin}\sim \rho r^3 v^2$ and the surface energy is $E_{\rm surf}\sim \gamma r^2$, where $\rho$ is the fluid density and $\gamma $ the surface tension. The relative velocity is expected to be proportional to the size of the droplet: $v\sim r/\tau$, where $1/\tau $ is the rate of shear of the flow which generates the droplets. The flow of liquid material is expected to be in turbulent motion, and the timescale $\tau $ can be identified with the Kolmogorov microscale of the turbulence, $\tau=\sqrt{\nu/{\cal E}}$, where $\nu$ is the kinematic viscosity of the fluid, and ${\cal E}$ is the rate of dissipation per unit mass. Equating the kinetic and surface energy scales, we can determine the approximate size of the droplets:
\begin{equation}
\label{eq: 5.1}
r\sim \left(\frac{\gamma \nu}{\rho {\cal E}}\right)^{1/3}\ .
\end{equation}
Note that because of the small exponent, the predicted size of relatively insensitive to the choice of parameters. We estimate the rate of dissipation per unit mass as ${\cal E}\sim v^2_{\rm col}/t_{\rm col}$, where $v_{\rm col}$ is the velocity with which the planets are colliding, and $t_{\rm col}\sim a/v_{\rm col}$ is the time required for the collision process, $a$ being the size of the smaller planet. To make a numerical estimate, we set $v_{\rm col}=10^3\,{\rm ms}^{-1}$, $a=10^7\,{\rm m}$, so that ${\cal E}\sim 100\,{\rm m}^2{\rm s}^{-3}$. We use the values $\rho\sim 3\times 10^3\,{\rm kg}\,{\rm m}^{-3}$, $\gamma \sim 3\times 10^{-2}\,{\rm Nm}^{-1}$ (a value typical for covalent liquids), $\nu \sim 10^{-2}\,{\rm m}^2{\rm s}^{-1}$ (a typical kinematic viscosity of glasses at extremely high temperatures). With these numerical values, equation (\ref{eq: 5.1}) gives sizes of order $r\sim 10^{-3}\,{\rm m}$, which is a typical size for chondrules.

\section{Concluding remarks}
\label{sec: 6}

We have argued for a hypothesis of planet formation by direct gravitational collapse, which is radically different from the standard model. Our hypothesis avoids the difficulties of explaining how dust particles can aggregate in a highly turbulent environment and of how metre-sized objects can avoid spiralling in to the star. It also suggests explanations of phenomena which do not have a satisfying explanation within the framework of the standard model.

Some quantitative predictions are in satisfactory agreement with observations. We showed that, depending upon initial conditions, the model allows for planets to reach circular orbits, as well as remaining in eccentric orbits. We argued that the FU Orionis outbursts could result from collisions, and showed that our model is consistent with a small number of collisions between juvenile planets over the lifetime of the disc, which is consistent with the observed frequency of outbursts. Our hypothesis also gives reasonable estimates for the timescales of rise and decay of an FU Orionis outburst, and of its magnitude. Also, one of the possible routes to the formation of chondrules within our model gives results which are consistent with the typical size of chondrules.

There is considerable scope for further work to determine the extent to which our arguments are supported by more detailed and quantitative investigations. Also because of the complexity of the processes involved in transforming our hypothesised juvenile planets into a mature planetary system, there may be other mechanisms which we have not envisaged. Our model raises new questions which are likely to be a fertile ground for subsequent research.

\section*{Acknowledgments}

Support from Vetenskapsr\aa{}det is gratefully acknowledged.

\end{document}